\documentstyle[12pt,epsfig]{article}
\advance\textheight by \topskip
\begin{document}
\tolerance=100000
\thispagestyle{empty}
\setcounter{page}{0}

\def\mathrm{\rm}
\newcommand{\be}{\begin{equation}}
\newcommand{\ee}{\end{equation}}
\newcommand{\br}{\begin{eqnarray}}
\newcommand{\er}{\end{eqnarray}}
\newcommand{\ba}{\begin{array}}
\newcommand{\ea}{\end{array}}
\newcommand{\bi}{\begin{itemize}}
\newcommand{\ei}{\end{itemize}}
\newcommand{\bn}{\begin{enumerate}}
\newcommand{\en}{\end{enumerate}}
\newcommand{\bc}{\begin{center}}
\newcommand{\ec}{\end{center}}
\newcommand{\ul}{\underline}
\newcommand{\ol}{\overline}
\newcommand{\ar}{\rightarrow}
\newcommand{\sm}{${\cal {SM}}$}
\newcommand{\susy}{{{SUSY}}}
\newcommand{\Dir}{\kern -6.4pt\Big{/}}
\newcommand{\Dirin}{\kern -10.4pt\Big{/}\kern 4.4pt}
\newcommand{\DDir}{\kern -10.6pt\Big{/}}
\newcommand{\DGir}{\kern -6.0pt\Big{/}}
\def\epem{\ifmmode{e^+ e^-} \else{$e^+ e^-$} \fi}
\newcommand{\eeWW}{\ifmmode{e^+e^-\rightarrow W^+ W^-} 
                   \else{$e^+e^-\rightarrow W^+ W^-$}\fi}
\newcommand{\eeHH}{\ifmmode{e^+e^-\rightarrow H^+ H^-} 
                   \else{$e^+e^-\rightarrow H^+ H^-$}\fi}
\newcommand{\WW}{\ifmmode{W^+W^-}\else{$W^+W^-$}\fi}
\newcommand{\HH}{\ifmmode{H^+H^-}\else{$H^+H^-$}\fi}
\newcommand{\csnt}{\ifmmode{c\bar s\tau \bar\nu_\tau}
                   \else{$c\bar s\tau \bar\nu_\tau$}\fi}
\newcommand{\cs}{\ifmmode{c\bar s}
                   \else{$c\bar s}\fi}
\newcommand{\nt}{\ifmmode{\tau \bar\nu_\tau}
                   \else{$\tau \bar\nu_\tau$}\fi}
\def\sm{\ifmmode{{\cal {SM}}}\else{${\cal {SM}}$}\fi}
\def\mssm{\ifmmode{{\cal {MSSM}}}\else{${\cal {MSSM}}$}\fi}
\def\MZ{\ifmmode{{M_{Z}}}\else{${M_{Z}}$}\fi}
\def\MH{\ifmmode{{M_{H}}}\else{${M_{H}}$}\fi}
\def\Mh{\ifmmode{{M_{h}}}\else{${M_{h}}$}\fi}
\def\MA{\ifmmode{{M_{A}}}\else{${M_{A}}$}\fi}
\def\MHpm{\ifmmode{{M_{H^\pm}}}\else{${M_{H^\pm}}$}\fi}
\def\MWpm{\ifmmode{{M_{W^\pm}}}\else{${M_{W^\pm}}$}\fi}
\def\tb{\ifmmode{\tan\beta}\else{$\tan\beta$}\fi}
\def\ctb{\ifmmode{\cot\beta}\else{$\cot\beta$}\fi}
\def\ta{\ifmmode{\tan\alpha}\else{$\tan\alpha$}\fi}
\def\cta{\ifmmode{\cot\alpha}\else{$\cot\alpha$}\fi}
\def\tba{\ifmmode{\tan\beta=1.5}\else{$\tan\beta=1.5$}\fi}
\def\tbb{\ifmmode{\tan\beta=30}\else{$\tan\beta=30$}\fi}
\def\cab{\ifmmode{c_{\alpha\beta}}\else{$c_{\alpha\beta}$}\fi}
\def\sab{\ifmmode{s_{\alpha\beta}}\else{$s_{\alpha\beta}$}\fi}
\def\cba{\ifmmode{c_{\beta\alpha}}\else{$c_{\beta\alpha}$}\fi}
\def\sba{\ifmmode{s_{\beta\alpha}}\else{$s_{\beta\alpha}$}\fi}
\def\ca{\ifmmode{c_{\alpha}}\else{$c_{\alpha}$}\fi}
\def\sa{\ifmmode{s_{\alpha}}\else{$s_{\alpha}$}\fi}
\def\cb{\ifmmode{c_{\beta}}\else{$c_{\beta}$}\fi}
\def\sb{\ifmmode{s_{\beta}}\else{$s_{\beta}$}\fi}
\def\Ord{\buildrel{\scriptscriptstyle <}\over{\scriptscriptstyle\sim}}
\def\OOrd{\buildrel{\scriptscriptstyle >}\over{\scriptscriptstyle\sim}}
\def\pl #1 #2 #3 {{\it Phys.~Lett.} {\bf#1} (#2) #3}
\def\np #1 #2 #3 {{\it Nucl.~Phys.} {\bf#1} (#2) #3}
\def\zp #1 #2 #3 {{\it Z.~Phys.} {\bf#1} (#2) #3}
\def\pr #1 #2 #3 {{\it Phys.~Rev.} {\bf#1} (#2) #3}
\def\prep #1 #2 #3 {{\it Phys.~Rep.} {\bf#1} (#2) #3}
\def\prl #1 #2 #3 {{\it Phys.~Rev.~Lett.} {\bf#1} (#2) #3}
\def\mpl #1 #2 #3 {{\it Mod.~Phys.~Lett.} {\bf#1} (#2) #3}
\def\rmp #1 #2 #3 {{\it Rev. Mod. Phys.} {\bf#1} (#2) #3}
\def\sjnp #1 #2 #3 {{\it Sov. J. Nucl. Phys.} {\bf#1} (#2) #3}
\def\cpc #1 #2 #3 {{\it Comp. Phys. Comm.} {\bf#1} (#2) #3}
\def\xx #1 #2 #3 {{\bf#1}, (#2) #3}
\def\preprint{{\it preprint}}

\begin{flushright}
{Cavendish-HEP-96/15}\\ 
{DFTT 49/96}\\ 
{November 1996\hspace*{.5 truecm}}\\
\end{flushright}

\vspace*{\fill}

\begin{center}
{\Large \bf 
$e^+e^-\ar H^+H^-$ signals at LEP2 energies\\
in the Minimal Supersymmetric Standard Model}\\[1.5cm]
{\large Stefano Moretti$^{a,b}$ and Kosuke Odagiri$^{a}$}\\[0.4 cm]
{\it a) Cavendish Laboratory, University of Cambridge,}\\
{\it Madingley Road, Cambridge CB3 0HE, UK.}\\[0.25cm]
{\it b) Dipartimento di Fisica Teorica, Universit\`a di Torino,}\\
{\it and I.N.F.N., Sezione di Torino,}\\
{\it Via Pietro Giuria 1, 10125 Torino, Italy.}\\[0.5cm]
\end{center}
\vspace*{\fill}

\begin{abstract}
{\noindent 
\small 
In this paper we compare $H^+H^-$ and $W^+W^-$ into four-fermion 
production at centre-of-mass energies typical of LEP2 and 
somewhat larger. The theoretical framework considered
is the Minimal Supersymmetric Standard Model.
The interest in exploiting the $e^+e^-$ CERN collider at values of $\sqrt s$
greater than 192 GeV  could come from the discovery of 
Supersymmetric 
signals during runs at lower energy. If these indicate that a
charged Higgs boson exists in the mass range $\MH\approx95-105$ GeV,
then a few years of running at $\sqrt s=205-215$ GeV and nominal luminosity
could make the detection of such scalars feasible, in 
the purely leptonic channel $\tau\nu_\tau\tau\nu_\tau$
and, for small $\tb$'s, also in
the semi-hadronic(leptonic) one ${\mathrm{jj}}
\tau\nu_\tau$. Charged Higgs bosons of the above nature 
cannot be produced by the beam energies approved at present for LEP2.
However, if runs beyond the
so-called `192 GeV cryogenic limit' will be approved by the CERN Council,
our selection procedure will enable us to establish the presence,
or otherwise, of charged Higgs bosons in the mentioned mass range.}
\end{abstract}
\vskip1.0cm
\hrule
\vskip0.25cm
\noindent
Electronic mails: moretti,odagiri@hep.phy.cam.ac.uk

\vspace*{\fill}
\newpage

\section*{1. Introduction}

The LEP2 collider \cite{lep2w} has already started operations. 
Event samples have been collected so far at $\sqrt s\approx130$ and 161 GeV 
as well as at $\sqrt s\approx172$ GeV. Further runs are scheduled at 
192 GeV. Although not yet approved, the value
$\sqrt s=205$ GeV has also been considered in the context of the 1995
LEP2 Workshop \cite{lep2w}. 
From the results reported so far by the LEP Collaborations,
it can be inferred that at present, 
contrary to the first excitement (which 
led many to believe in the evidence of some physics in the Supersymmetric
sector), the data from the runs above
the $Z$ peak are in fair agreement with
the Standard Model (\sm) expectations \cite{lep1.5}\footnote{Although
it should be mentioned that the ALEPH Collaboration
still claims to see in its data a rather significant excess of four-jet events
in the sum of the di-jet masses at $\approx105$ GeV, at all LEP2
energies considered so far \cite{ragusa}. In contrast, the other Collaborations
have never established the presence of similar effects.}.

However, it is one of the primary goals of the LEP2 CERN collider to possibly
bring some new insight to clarify whether already at its energy scale
physics beyond the \sm\ exists or not,
particularly in the Higgs sector of the electroweak
(EW) interactions \cite{Higgs}. In this respect, one of the most appealing
theoretical framework is the Minimal Supersymmetric Standard Model (\mssm).
In fact, being a Supersymmetric (SUSY) theory, it provides a solution to the 
hierarchy problem via the cancellations of the infinities arising
in the \sm\ when computing quartic corrections to the Higgs mass and, at
the same time, these can be kept under control if SUSY breaking occurs
sufficiently near the electroweak scale $G_F^{-1/2}$. Moreover, if it does so 
and if it is combined with Grand Unification Theories, the \mssm\ predicts
a value for the Weinberg angle $\theta_W$ in good agreement with the measured
one and a value for the Grand Unification Mass that can
explain the non-observed proton decay \cite{protondecay}. It also supplies a
natural candidate for the dark matter in terms of the Lightest Supersymmetric
Particle ({${\cal {LSP}}$), which is stable, neutral and weakly
interacting (the neutralino).

The structure of the Higgs sector of the \mssm\ is determined at 
tree-level
by only two 
parameters\footnote{At one-loop also the top mass and the squark masses
are needed. For these, we assume throughout the paper: $m_t=175$ GeV and
$M_S=1$ TeV.}: for example, $\MA$ (the mass of the pseudoscalar
Higgs) and $\tb$ (the ratio between the vacuum expectation values
associated with the two Higgs doublets, one giving masses to the 
up-type and the other to the down-type fermions). Thus, such a model
is also rather predictive, not much less than the \sm, for which one
parameter (the Higgs mass $M_\phi$) 
is needed to describe its phenomenology.
Other than the pseudoscalar neutral Higgs $A$, the \mssm\ Higgs spectrum
involves other four scalars: the ${\cal {CP}}$--even neutral $H$ and $h$,
and the two charged  $H^\pm$'s.

The search for Higgs bosons of the \mssm\ at LEP2 will be based mainly
on the Higgs-strahlung process $e^+e^-\ar Zh$ and on the associated
production $e^+e^-\ar Ah$ \cite{wrk39}. These two mechanisms are somehow
complementary. In fact, for small values of $\tb$ the first process dominates,
whereas at large $\tb$'s the second reaction becomes quantitatively 
more important \cite{Higgs} (provided it is kinematically allowed). 
Typical cross sections are of the order 
of one picobarn or so. The dominant decay modes of both the $h$ and $A$
scalars are into $b\bar b$ and $\tau^+\tau^-$ pairs\footnote{In certain areas
of the Supersymmetric parameter space Higgs particles can also
decay into the SUSY-partners of ordinary particles, such as into
invisible $\chi^0_1\chi^0_1$ (${\cal {LSP}}$) pairs or via other 
modes involving charginos and/or neutralinos \cite{wrk40,wrk41}.} 
\cite{ioejames}.
Concerning the heavy neutral Higgs
$H$, it should be noted that its mass (being much larger than the
$Z$ mass) implies that the chances for $H$ production 
are confined to a tiny corner of the $(\MA,\tb)$ parameter
space (i.e., very small $\MA$ and $\tb$), where the cross
section is very small and also P-wave suppressed at threshold
\cite{Higgs}.

The situation for the charged Higgs $H^\pm$ is apparently even less 
optimistic and extremely complicated. Indeed, the cross section for
charged Higgs boson production via the channel 
$e^+e^-\ar H^+H^-$ \cite{wrk41,wrk112}
(although not particularly small at LEP2) yields a signal which
is very hard to extract, because of the huge irreducible background
in $e^+e^-\ar W^+W^-$
events. In fact, on the one hand, the \mssm\ mass relations tell us 
(at tree-level) that $\MHpm^2=\MWpm^2+\MA^2$ and, on the other hand,
kinematic bounds dictated by the LEP2 centre-of-mass (CM) 
energy imply that only $H^\pm$ scalars with mass $\MHpm\Ord\sqrt s/2$
can be produced.
Since the highest CM energy 
run scheduled for the CERN machine is at $\sqrt s=192$ GeV, this realistically
means that only charged Higgses with $\MHpm\Ord90$ GeV can 
have a sizable cross section at LEP2. Having at present
the lower limit on the \mssm\ charged Higgs mass set
at 85 GeV, this leaves only
a small window at LEP2. Furthermore, 
in the above region, the $H^\pm$ and $W^\pm$ bosons are almost
degenerate in mass, such that the differential distributions involving their
decay products can hardly be useful in disentangling the two signals, 
especially considering that at $\sqrt s\approx200$ 
GeV the cross section for $W^+W^-$
production is almost three orders of magnitude larger
than that for $H^+H^-$\footnote{Note that
the mentioned mass relation does not hold in non-minimal (NM) SUSY
extensions of the \sm, for which the only lower mass constraint 
comes from LEP1 data, $M_{H^\pm}^{\mathrm{NM}}\OOrd44$ GeV \cite{Higgs}.
In this context, LEP2 has a good potential for the charged Higgs boson
discovery, especially for $M_{H^\pm}<\MWpm$ \cite{Higgs}.} !
The typical signature of a $H^\pm$ scalar would be most likely
an excess of
$\tau$ events with respect to the rates predicted by the \sm\
({\sl lepton universality breaking} signal), as the lepton-neutrino
decay channel has the largest branching ratio (BR). For small $\tan\beta$'s,
hadronic channels can be also considered \cite{Higgs}, together with 
the analysis of `cascade' decays $H^\pm\ar W^{\pm *}h,W^{\pm *}A$ 
(with $h,A\ar b\bar b$). 

It is the purpose of this study to reanalyse the discovery potential
of $H^\pm$ particles of the \mssm\ at the LEP2 machine at CM energies of
not only 205 GeV but also higher, 
for example: $\sqrt s=210$ and 215 GeV. 
We stress that increasing $\sqrt s$ would generally bring some advantages in 
the above respect, and that a 10 GeV difference in $\sqrt s$ could be 
crucial to detecting charged Higgses of masses up to 100 GeV or so.
\begin{itemize} 
\item First, the window in $\MHpm$ that could be explored would be larger.
\item Second, if a $H^\pm$ scalar with a mass significantly larger
than that of the $W^\pm$ (say,  by 20 GeV) can be produced, 
then the different kinematics of
their decay products (jet and leptons) can be fully exploited.
In practise, particles 
arising from the $H^\pm$'s would have in the collider CM the same kinematics
as in the decay frames of the  scalars, whereas those
coming from the $W^\pm$'s would have a significant boost.
\item Third, even compared to the rates at 205 GeV,
increasing $\sqrt s$ up to 215 GeV would enhance the cross section
of $H^+H^-$ production by a factor of four (for $\MHpm\approx100$ GeV),
whereas that for the creation of $W^+W^-$ pairs
would remain roughly constant. 
\end{itemize}
The motivations for considering a run at a CM energy larger than 
the highest one scheduled at present for LEP2, in view of a possible
$H^\pm$ detection, could be justified by
two scenarios that could well occur at previous energy stages.
\begin{enumerate}
\item One or more Higgs bosons are discovered and these match the
\mssm\ particle spectrum (i.e., a $h$ or/and an $A$ scalar).
In addition, the corresponding masses are so light that they are 
consistent with a value of $\MHpm$ just beyond the reach of
$\sqrt s=192$ GeV.
For example, note that, for $\MA\approx60$ GeV,
the \mssm\ predicts $\Mh\approx55-60~{\mathrm{ GeV}}$ 
and $\MHpm\approx100$ GeV, but
the corresponding $H^+H^-$
cross section is too small even at 205 GeV 
\cite{Higgs}.
%
%
%
\item Some other \mssm\ signal is detected, and this is compatible
with a charged Higgs boson around $100$ GeV. For example, for
$h$ and $A$ masses below 70 GeV (corresponding to $\MHpm\Ord106$ GeV)
and for relatively small $\tb$, there are still $(\MA,\tb)$ open windows 
for $h,A\ar$ neutralinos decays \cite{wrk41,wrk115}.
\end{enumerate}
If one of these two cases occurs, then it might be conceivable
to put some more effort to increasing the discovery potential of a machine
already existing, rather than to wait more years 
until the same Higgs mass domain can be covered by other machines
(in particular, the LHC and the NLC). 

The possibility of running the LEP2 machine at CM energies of
205 GeV or more has been already studied in the context of the LEP2 
Workshop \cite{upgrade} (see also Ref.~\cite{chamonix}), even though 
the maximum CM energy stage approved to date by the CERN Council is 192 GeV. 
(In particular, Ref.~\cite{upgrade} considered beam energies up to 104 GeV.)
Indeed, a beam energy of 96 GeV (Phase IV) represents a clear `turning point' 
in the `LEP Energy Upgrade Programme', as beam energies larger than 
that would require an upgrade of the cryoplants 
(Phases X and Y) now in use at LEP, 
because of their limited cooling power. 
In this respect, however, we would like to stress that the three values
$\sqrt s =205,210$ and 215 GeV are {\sl all} 
beyond the present so-called `cryogenic limit', such that they would require 
the same kind of layout modifications and equipment fittings. 
Furthermore, we remind
the reader the following statement in the LEP2 Proceedings
(incidentally, precisely where the \sm\ and the \mssm\ Higgs sectors
are compared to each other): ``{\sl It could
however be imagined that if a new particle were discovered, a substantial
extension of the LEP2 project could be decided upon with the purpose 
of identifying this particle}'' \cite{citation}. We push our attitude
even further, by saying that if this particle is a \mssm\ one and this
tell us that some more \mssm\ physics is just {\sl beyond the corner}, such
an {\sl extension} could also well be in the direction of a further
beam energy upgrade.
In contrast, if none of the above options (i.e., 1. and 2.) is verified, 
the prospect 
of LEP2 running at 205 GeV or above that would have no special meaning from 
our viewpoint.

The plan of the paper is as follows. In Section 2 we describe our
computational technique and give the values of the parameters used. 
In Section 3 we present and discuss our
results. In Section 4 we draw our conclusions.

\section*{2. Calculation} 

Many analytical formulae exist to compute the total cross section 
for on-shell $H^+H^-$ production \cite{wrk112,ee500}. 
In contrast, the literature on the
off-shell production, followed by the decays of the two bosons (for example,
into four fermions) is more limited. Thus, in carrying out our analysis
we have decided to recompute the matrix element for the $2\ar4$ process
\be\label{eeHH}
e^+e^-\ar H^+H^-\ar f_1\bar f_2 f_3\bar f_4,
\ee
where $f_{1,2,3,4}$ represent generic fermions.
To this end we have proceeded in the following way. First, 
we have used the helicity amplitude techniques 
described in Refs.~\cite{KS,mana,ioPR}
to produce a {\tt FORTRAN} code computing the Feynman amplitude squared
of process (\ref{eeHH}). Furthermore, this has been counter-checked against
a second program built up by using the subroutines contained in the
package {\tt HELAS} \cite{HELAS}, which indeed include 
scalar-fermion vertices (for the Goldstone bosons of the \sm), with
the couplings rearranged to the values predicted by the \mssm.
Finally, a third program implementing formulae obtained by using
the textbook method of tracing the $\gamma$-matrices was also generated.
The results obtained with the three different formalisms agree within 
twelve figures in {\tt REAL*8} precision. The matrix elements have been then
integrated over the phase space and the distributions produced by 
resorting to a multidimensional integration making use of the Monte Carlo (MC)
routine {\tt VEGAS} \cite{VEGAS}.
Also to compute the rates  for the \sm\ reaction
\be\label{eeWW}
e^+e^-\ar W^+W^-\ar f_1\bar f_2 f_3\bar f_4
\ee
we have proceeded as above. We are aware that a huge number of analytical, 
semi-analytical and numerical calculations are available in literature
for reaction (2)
\cite{4fermions}. However, to our purposes, it was more useful to have a
handy program, constructed on the same footing as that for 
four-fermion production via $H^+H^-$, 
with the advantage of offering
a straightforward comparison between the rates
of the two mechanisms (\ref{eeHH})--(\ref{eeWW}). 

Concerning the possible signatures of process (\ref{eeHH}) and (\ref{eeWW}),
one can expect 
\be\label{jjjj}
H^+H^-, W^+W^-\ar {\mathrm{jjjj}},\quad\quad\quad{\mathrm{hadronic~channel}},
\ee
\be\label{jjln}
H^+H^-, W^+W^-\ar {\mathrm{jj}}
\tau\nu_\tau,\quad\quad\quad{\mathrm{semi-hadronic(leptonic)~channel}},
\ee
\be\label{lnln}
H^+H^-, W^+W^-\ar \tau\nu_\tau
\tau\nu_\tau,\quad\quad\quad{\mathrm{leptonic~channel}}.
\ee
In the following, we will consider all of them.

A few details now, concerning the numerical values of the parameters
entering in the calculation.  The numbers adopted are: 
$\alpha_{em}= 1/128$, $\sin^2\theta_W=0.2320$, 
$M_{Z}=91.1$ GeV, $\Gamma_{Z}=2.5$ GeV,
$M_{W^\pm}\equiv M_Z\cos\theta_W \approx 80$ GeV and
$\Gamma_{W^\pm}=2.2$ GeV, while for the fermion masses we have used
$m_\tau=1.78$ GeV, $m_{\nu_\tau}=m_u=m_d=0$, $m_s=0.3$ GeV
and $m_c =1.4$ GeV. The $H^\pm$ mass has been computed using the 
tree-level relation $M_{H^\pm}^2=M_{A}^2+M_{W^\pm}^2$,
as one-loop corrections are small \cite{corrMHMSSM}.
The $H^\pm$ width has been taken from 
Ref.~\cite{ioejames}. The strong coupling constant 
$\alpha_s$, which enters in the {\sl running masses} of the quarks
(for details, see Ref.~\cite{ioejames}) when calculating 
$H^\pm\ar cs$ decays, has been
evaluated
at the two loop order, with $\Lambda^{(4)}_{\overline{\mathrm {MS}}}=230$ MeV,
at the scale $Q^2=s$.

To allow for large rates in the hadronic and semi-hadronic(leptonic) 
Higgs decay
channels, we have restricted our attention to the case of small
values of $\tan\beta$, since
for large values the $H\ar cs$ channel becomes negligible.   
For reference, in presenting our results we will adopt the choice 
$\tba$\footnote{Note that the $\tb$ dependence of process (\ref{eeHH})
enters when considering Higgs decays, since the rates
of the on-shell production depend only on $\MHpm$.}. 
Finally, for the mass of the charged Higgs boson we will use
the value $\MHpm\approx100$ GeV (that is, $\MA=60$ GeV).
However, we will generalise in the end our conclusions also to other
values of the two \mssm\ parameters. 

A brief comment is in order before ending this section, concerning
higher order corrections to processes (\ref{eeHH})--(\ref{eeWW}).
Many of these are well known in literature, in both cases 
\cite{corrHH,corrWW}. However, we have not included them in our
analysis. In fact, on the one hand, we are interested in 
establishing effects (differences in total and differential rates
between the two processes) whose relative magnitude must be well beyond
the effects due to radiative corrections, in order to be detectable.
On the other hand, higher order results
can be easily incorporated (in the case of process (1))
or are already included (in the case
of process (2)) in many of the calculations available in literature.
In the very end, in fact, our results should be reproduced
by the various event generators \cite{events} that are in possess
of the four
LEP2 Collaborations and which include all the above corrections, together
with a simulation of the various detector and hadronisation effects,
which 
are clearly beyond our capabilities.

\section*{3. Results}

For the values that we have assumed for $\tb$,  $\MHpm$
and $\MWpm$, the total cross
sections for processes (1)--(2) are displayed in the
upper sections of Tab.~Ia, b and c (in correspondence to the 
three mentioned decay channels). The values of energies considered
are $\sqrt s=205,210$ and 215 GeV.
From those numbers it is clear that, if no appropriate Higgs selection 
procedure is exploited, the chances of disentangling
$H^+H^-$ signals from the enormous $W^+W^-$ backgrounds are practically 
nil: and this is true for all decay channels 
(\ref{jjjj})--(\ref{lnln})
and CM energies as well.
Differences are dramatic in the four-jet channel, but these gradually
diminish if the lepton-neutrino decays are selected. Nevertheless, at this
stage, the $H^+H^-$ rates are at the most $12\%$ of the $W^+W^-$ ones
(this occurs at $\sqrt s=215$ GeV for the leptonic signature).
This is too little, further considering that in the 
$\tau\nu_\tau\tau\nu_\tau$ channel invariant mass spectra for the 
decay products of the two bosons cannot be reconstructed because
of the presence of {\sl two neutrinos}. 

The relative statistical importance 
of the rates in Tab.~I for the three channels (3)--(5) 
can be understood in term of the corresponding branching ratios 
(BRs) (see, e.g., Ref.~\cite{ioejames})\footnote{Assuming 
again the mentioned values of $\tb$ and 
$\MHpm$.}:
\be\label{BRjjjj}
BR(H^+H^-\ar {\mathrm{jjjj}})\approx1\%,
\quad\quad\quad 
BR(W^+W^-\ar {\mathrm{jjjj}})\approx50\%,
\ee
\be\label{BRjjln}
BR(H^+H^-\ar {\mathrm{jj}}\tau\nu_\tau)\approx20\%,
\quad\quad\quad 
BR(W^+W^-\ar {\mathrm{jj}}\tau\nu_\tau)\approx14\%,
\ee
\be\label{BRlnln}
BR(H^+H^-\ar \tau\nu_\tau\tau\nu_\tau)\approx63\%,
\quad\quad\quad 
BR(W^+W^-\ar \tau\nu_\tau\tau\nu_\tau)\approx1\%.
\ee
In general, it has to be said that leptonic decays favour
$H^\pm$ signals, whereas hadronic decays favour
$W^\pm$ backgrounds, this behaviour being dictated by the interplay 
between the values of
the fermion masses and $\tan\beta$ in the scalar-fermion-fermion
couplings. In fact, one gets 
that $BR(H^\pm\ar \tau\nu_\tau)/BR(W^\pm\ar\tau\nu_\tau)\approx7$ and
$BR(W^\pm\ar {\mathrm{jj}})/BR(H^\pm\ar{\mathrm{jj}})\approx6$.

As mentioned in the Introduction, there are systematical differences
in the kinematics of events of the type (1)--(2) that could help in
extracting the signals out of the backgrounds. 
\begin{enumerate}
\item $H^+H^-$ production proceeds via the $s$-channel only, whereas 
$W^+W^-$ events can occur through $t$- and $u$-channels via the
exchange of an electron neutrino. (i) As a consequence, in the latter case,
one would expect to see in the spectrum of the angle of the $W^\pm$ boson(s)
with respect to the beam a component
peaking in the forward and/or backward direction. Indeed, it turns
out that at LEP2 energies the contribution of the graph
involving neutrino exchange is  larger (by a factor of 2)
than that of the graphs involving the triple gluon vertices $\gamma W^+W^-$
and $Z W^+W^-$
(for details, see Ref.~\cite{anomalous}).
In contrast, in the former case,
the angular distribution of the charged Higgs bosons
should follow the $\sin^2\theta$ law typical of spin-zero particle
production.
\item Since the $H^\pm$ bosons are produced via process (1) practically at
rest, the kinematics of their decay products only depends
on the actual value of $\MHpm$ and retains most of the features that it
would have in the $H^\pm$ rest frame. In contrast, the $W^\pm$ decay 
products suffer from a boost, dictated by the value of the difference
$\sqrt s-2\MWpm$. Hence, one expects that, for example:
(ii) the two fermions coming from the same $H^\pm$ decay are back-to-back
in the collider CM frame;
(iii) their transverse momentum has a low maximum value.
These features are not shared by $W^+W^-$ events. 
\end{enumerate}
%
We notice 
that only in one of the three cases (3)--(5) the above quantities
can be defined unambiguously. This happens for
the semi-hadronic(leptonic) channel. In the cases of the hadronic and
leptonic signatures this is no longer true.
In the former case, 
in only one of the three possible two-jet combinations the particles
would come from the same decay.
In the latter case, the presence of two neutrinos prevents one even from
reconstructing the four-vectors of the particles that decay, as only three
four-momenta can be assigned (one for each of
the $\tau$'s and the one defined by using the missing energy and 
three-momentum).
However, it is possible to remedy these two problems. Since the 
optimal solution depends on the signature considered, we treat separately 
the decays (3)--(5) in the three following subsections.
Moreover, in the forthcoming Figs.~1--4, the cases (a), (b) and
(c) will refer to the mentioned decays, in corresponding order, as
in Tab.~I\footnote{This has been done in order to facilitate the comparison
among the various decay channels, at the cost of discussing
the various figures in a different order respect to that in which they 
are plotted. Note also that distributions in Figs.~1--3 are normalised to
unity, whereas those in Fig.~4 sum to the total cross sections.}.

\subsection*{3.1 The hadronic channel}

In the case of the decays (\ref{jjjj}) there is an effective
strategy to adopt, in order to assign the right pair of jets to the 
parent $H^\pm$:
that is, to assume that the two quarks belonging to the same boson decay
are those most far apart. In practise then, 
one can compute all the
relative angles between the directions of the four jets and couple
the two partons for which such an angle is largest.
 In contrast, the direction of any of the other
two quarks from the other $H^\pm$ decay is distributed in a 
random way with respect to the previous two, such that
only in a small corner of the phase space wrong assignments can occur. 
Naturally, what has just been said has also been done for $W^+W^-$
events. In this case, however, wrong assignments will be more frequent,
because the boosts on the decay products tend to shrink
their relative angle. 

Indeed, by looking at Fig.~1a, 2a and 3a, one realises that
the correct assignment is made most of the times, as the
variables plotted there largely retain the features that they would
have if the tagging was straightforward\footnote{Compare to 
Figs.~1b, 2b and 3b for the semi-hadronic(leptonic) decays
(see later on).}.
In detail, the angle of the reconstructed boson (i.e., $\theta_{\mathrm{B}}$)
follows to a large extent the expected $\sin^2\theta_{\mathrm{B}}$ (for
$H^+H^-$) and $\cos^2\theta_{\mathrm{B}}$ (for $W^+W^-$) laws (Fig.~1a).
At the same time, the relative angle of the two reconstructed jets
(i.e., $\theta_{\mathrm{ff}}$) belonging to the same $H^\pm$ decay 
tends to be distributed at small cosines, whereas for those from
the $W^\pm$ is much more spread out.
Finally, the spectrum in transverse momentum of the two jets 
(i.e., $p_T$) is much softer in the former than in the latter case.

It is then clear from the figures
that a Higgs selection strategy appears evident.
For example, one can impose the following kinematic cuts: 
$|\cos\theta_{\mathrm{B}}|<0.60$ (for all $\sqrt s$), plus
$\cos\theta_{\mathrm{ff}}$$<-0.88(-0.80)[-0.72]$ 
and $p_{\mathrm{T}}<24(34)[40]$ GeV, 
corresponding to $\sqrt s=205(210)[215]$ GeV, to enhance the
signal content in four-jet events. Unfortunately,  this
does not help much in the end. In fact, although the background contributions
are cut out by a factor of $\approx24(14)11$ at 
$\sqrt s=205(210)[215]$ GeV, whereas the signal  reduction is only 
$\approx1.3$ (at all energies), the cross sections of $H^+H^-$ events
are too small to be
observable, around three orders of magnitude 
smaller than those for $W^+W^-$ pairs (lower section
of Tab.~Ia). This can be 
seen in Fig.~4a, which shows the invariant mass spectrum
of the two jets selected (i.e., $M_{\mathrm{ff}}$). The Higgs
peaks are lost beneath the distributions of $W^+W^-$ events,
which (incidentally) hardly show a peak at 80 GeV.

\subsection*{3.2 The semi-hadronic(leptonic) channel}

In the case of the semi-hadronic(leptonic) channel the assignment
of the right pair of fermions to the parent boson is straightforward.
Therefore, the differential spectra do not suffer from any distortion and
reproduce exactly the features expected for the quantities plotted in
Fig.~1b, 2b and 3b\footnote{Note that in Fig.~1b both the angle $\theta_+$ and
$\theta_-$ (that is of the $W^+$ and $W^-$ respect to, say, the
electron beam direction) are plotted, even though
their distinction is possible 
in case of semi-hadronic(leptonic) decays.}.

The Higgs selection procedure we adopt here is the same as was outlined in the
previous section, as the behaviour
of the various quantities are very similar in both cases.
Contrary to hadronic decays, the final results in case of
channel (4) are encouraging indeed. In fact, from Tab.~Ib (lower part),
one can notice that after the kinematic cuts, the rates of
$H^+H^-$ and $W^+W^-$ events are now comparable, for all energies. Moreover,
on the one hand, the selection procedure is here even more effective
than in the hadronic channel and, on the other hand, the absolute
numbers are much bigger than before. 
The cross section for  $W^+W^-$ has been reduced by a factor
$\approx1075(491)[336]$, whereas for $H^+H^-$ events the numbers are
again $\approx1.3$  (for all $\sqrt s$'s).

The invariant mass distributions of the two-jet pair (or equivalently,
of the $\tau\nu_\tau$ pair, after the missing energy and three-momentum 
have been assigned to the neutrino) for case (4) are plotted in Fig.~4b.
The bins there are 2 GeV wide. It is worth noticing that the curious
shape of $W^+W^-$ events is simply an effect of the cuts, with no
special physics meaning. In fact, the mass spectra before the cuts
have `shoulders' (where the rates steeply drop) exactly where
the `spurious peaks' (those on the right in Fig.~4b) 
appear after the
kinematic cuts. In fact, the latter strongly cut-off configurations around
the $W^\pm$ mass, but not that much in regions further away. In the end
(luckily enough), this
leaves like a `valley' precisely where the Higgs peaks clearly
stick out, at all energies (and they would
even if the actual mass resolution of the detectors will be much worse than
2 GeV: note the logarithmic scale in the figure).

\subsection*{3.3 The leptonic channel}

The leptonic channel is probably
the most complicated to select kinematically,
because of the missing four-momentum due to the neutrinos.
Nonetheless, the large decay rates of the Higgs scalars into
$\tau\nu_\tau$ pairs furnish an appealing starting point.
In this case it is no longer possible to reconstruct pairs
of particles from the same boson decay in any way. The only thing which can
reasonably be done is to define the various variables by using the only two
four-momenta that can be tagged. Therefore, in this case: 
$\theta_{\mathrm{B}}$ is the angle of the system formed
by the two $\tau$'s respect to the beam directions;
$\theta_{\mathrm{ff}}$ is the angle between them;
$p_T$ their transverse momentum and $M_{\mathrm{ff}}$
their invariant mass.

No very distinctive features between the $H^+H^-$ and $W^+W^-$ decays 
appear in the spectrum of Fig.~1c, whereas some differences occur in
Fig.~2c and 3c. 
The first plot somewhat resembles those of the channels (3)--(4), at
least qualitatively.
In fact, there is a sort of kinematic symmetry in the two decays
of the bosons, with respect to each other, such that many of the
phase space configurations occupied by a neutrino in one decay
are also common to a lepton in the other one. 
Fig.~2c tells us that, in $W^+W^-$ events, whichever is the orientation
of the $\tau$ momenta, these tend to be aligned back-to-back by the
boosts, as the two $W^\pm$'s fly apart after 
they are produced.
In case of $H^+H^-$ decays, where no boost acts, the distribution is flat.
Furthermore, spin effects also act in the same direction.
The shape of the curves in Fig.~3c follows from that in Fig.~1c.
In particular, the tendency in $W^+W^-$ samples
of $|\cos\theta_{\mathrm{B}}|$ being closer to one implies that 
the corresponding 
$p_T$ spectrum (proportional to $|\sin\theta_{\mathrm{B}}|$)
is soft, whereas this does not occur for $H^+H^-$.

Having in mind the behaviours outlined in Fig.~1c, 2c and 3c,
we have attempted, as Higgs selection procedure, the one defined by the
following requirements on the phase space:
$\cos\theta_{\mathrm{ff}}$$>-0.50$ and $p_{\mathrm{T}}>50$ GeV, for 
$\sqrt s=205,210,215$ GeV 
(the spectrum in $\cos\theta_{\mathrm{B}}$ is no longer useful).
The rates that follow can be appreciated in Tab.~Ic (lower part).
Though much less than in the two previous cases, the signal-to-background
ratio has been improved.
In detail,
the cross section for  $W^+W^-$ has been reduced by a factor
$\approx5.4(5.6)[6.0]$, whereas for $H^+H^-$ events the numbers are
$\approx1.9(1.9)[1.9]$, corresponding to $\sqrt s=205(210)[215]$ GeV.
 
Finally, the invariant mass spectra of the $\tau^+\tau^-$ pair
is displayed in Fig.~4c. Since the energy of the leptons is more or less
the same both in $H^\pm$ and $W^\pm$ decays (on average, 1/4 of the
total CM energy) and since the $\tau$'s tend to be
back-to-back in $W^+W^-$ events (see Fig.~2c), 
their mass distributions are always shifted
towards high values, whereas this is not the case for the $H^+H^-$ sample
(for which the $\cos\theta_{\mathrm{ff}}$ distribution is flat).

\subsection*{3.4 Non-resonant backgrounds}

So far we have not mentioned that there are graphs that 
can produce the final states ${\mathrm{jjjj}},
{\mathrm{jj}}\tau\nu_\tau$ and $\tau\nu_\tau\tau\nu_\tau$ via $e^+e^-$
annihilations without proceeding  through $H^+H^-$ and $W^+W^-$ pair
production and decay. These can be EW contributions (for all three
final states) and QCD contributions as well (for the purely hadronic 
channel). In particular, the $ZZ$ resonant EW diagrams
(entering in the ${\mathrm{jjjj}}$ and $\tau\nu_\tau\tau\nu_\tau$
signatures) could well be large, as the CM energies considered here
are above the $ZZ$ threshold. In contrast, those involving 
$\gamma Z$, $\gamma\gamma$ intermediate states ($t$ and $u$ channel graphs)
as well as single resonant contributions ($s$ channel graphs) are always much 
smaller. QCD diagrams can be large, though their kinematics is very
different from that of the $H^+H^-$ and $W^+W^-$ resonant contributions.
Indeed, we expect the kinematic selections adopted here to largely reduce
the rates of background events.

We have calculated all these additional diagrams by using
the \cite{HELAS} libraries, and counter-checked (where possible) 
their outputs against the well-known results given in literature
\cite{4fermions}, obtaining perfect agreement. After the signal
selections adopted in this study and for the ${\mathrm{jj}}\tau\nu_\tau$
and $\tau\nu_\tau\tau\nu_\tau$ channels, 
the differences are at the level of percent or less, so that, e.g.,
including the new diagrams and interfering them with processes
(1) and (2) hardly modifies the distributions in Figs.~4a,b and c and
the number in the low sections of Tabs.~Ia--c. In particular, we
notice that the larger effects (up to $7\%$, at $\sqrt s=215$ GeV)
occur in the leptonic channel. However, these are confined in the low
invariant mass region of Fig.~4c, so that a further cut in that
region (e.g., $M_{\mathrm{ff}}>20$ GeV) allows one to remove completely
the background events, without affecting dramatically the 
$H^+H^-$ and $W^+W^-$ rates. In case of the ${\mathrm{jjjj}}$ channel
differences can be larger (especially depending on the selection criteria 
of the QCD events). However this is not relevant for our analysis as the
purely hadronic channel is useless to our purposes 
(see Fig.~4a and Tab.~Ia). 

\section*{4. Conclusions}

We conclude our paper in a rather optimistic way: 
in order to disentangle $H^\pm$ from $W^\pm$ decays (within the 
framework of the \mssm), a higher energy option of LEP2 
(beyond 200 GeV) could be helpful. 
In fact, on the one hand, it is certainly impossible to resort to
hadronic decays but, on the other hand, semi-hadronic(leptonic)
and leptonic decays would offer chances of detection. This is especially
true at 215 GeV and for the ${\mathrm{jj}}\tau\nu_\tau$ channel.
In this case, after an appropriate kinematic selection,
the integrated cross section 
of signal events is only approximately $23\%$ smaller than
that of background events. The number increases to $\approx33\%$ at
210 GeV and to $\approx40\%$ at 205 GeV. In this case, one can also
rely on the invariant mass spectra (of the pair
of fermions coming from the same boson decay) 
to disentangle $H^\pm$ peaks, which
should be clearly visible on top of the $W^\pm$ background.
For leptonic decays, the only way to assess the presence of charged
Higgs signals is to look for lepton universality breaking in $\tau$
production. In this case, one gets an excess of events (with respect
to the \sm\ predictions) of the order of 
$\approx8(21)[36]\%$, corresponding to the CM energies
$\sqrt s=205(210)[215]$ GeV. Furthermore, for the leptonic signature, 
some signal effects could also be visible in 
the distributions in the
invariant mass of the detectable $\tau^+\tau^-$ pair.

We also stress that changing the value of $\MHpm$ (in the range in which
Higgs signals are produced with large cross section, say between 
95 and 105 GeV, for $\sqrt s=215$ GeV) should not modify drastically the
above conclusions. In contrast, the actual value of $\tb$ is 
crucial, as the hadronic decay rates of the charged Higgs boson
strongly depend on this parameter (the larger this is the smaller they are),
especially because the semi-hadronic(leptonic) channel
is the most promising one. In fact, for large values of $\tb$ only
the fully leptonic signature would be exploitable. In this case, however,
the $H^+H^-\ar\tau\nu_\tau\tau\nu_\tau$ rates would be larger than those 
presented here. For example,
for $\tb=30$, the branching ratio of charged Higgses into $\tau\nu_\tau$-pairs
is practically $100\%$ (against $79\%$ at $\tb=1.5$), so that 
the $H^+H^-$ cross sections in Tab.~Ic would be increased in the end by a 
factor $\approx1.6$, inducing an excess of $\tau\nu_\tau\tau\nu_\tau$ events
of the order of $\approx13(33)[58]\%$. Also note that the key features of
our analysis are not exploitable for lighter $H^\pm$'s (those
producible at 192 GeV and below), simply because their mass would be 
{\sl degenerate}
with that of the $W^\pm$ and most of the quantities we resorted to
in the event selection 
would look identical in both the signal and the background.
Furthermore, we add that the inclusion of the Initial State Radiation
(ISR) should not modify the interplay between $H^+H^-$ and
$W^+W^-$ events, as in general it would not affect their relative rates.

The crucial point seems to be in the end the number of $H^+H^-$ events on which
one can rely. In the two detectable channels, the typical signal
cross sections are between 1 and 10 femtobarns
(they are bigger at higher CM energies).
Therefore, 
between 10 and 100 inverse picobarns would be needed
at 215 GeV to produce one signal per year. Since this is roughly the
luminosity which is expected to be collected at the various energy stages
(per experiment), we expect that after a few years of running at the higher
energy options considered here something like ten charged Higgs events of 
the \mssm\ could be detected. Finally,
before closing we would like to make two final considerations.

Firstly, we remind the reader that we have carried out our analysis, which
considers an increased LEP2 beam energy, 
under the assumption that this situation
is conceivable only if the presence of SUSY signals was already established
during runs at lower energies, with those indicating the existence of a charged
Higgs $H^\pm$ with mass around $100$ GeV 
(thus, beyond the reach of $\sqrt s=192$ GeV).
In other terms, one would be searching for something that he already
knows is existing. Therefore, the {\sl extension of the LEP2 project}
that we have mentioned at the beginning could well be also finalised in 
accumulating as much luminosity as needed to assess the presence
of the expected particle at the higher energies.

Secondly, we warn the reader that our results come from partonic 
calculations. No hadronisation or detector effects have been taken into
account. Not even Initial State Radiation was implemented.
Therefore before being taken as conclusive, these should be confirmed
by the experimental simulations, by means of MC event generators,
which include all the above aspects.
Even because, in doing so, the Higgs selection procedures that we have
adopted could also be improved, yielding more interesting arguments 
to the issue that we have raised here. 

In the end, in the case that further upgrades beyond the `192 GeV limit' 
will be eventually considered by the CERN Council, we believe that our results 
should be kept into account when planning the activity of the machine at
those new energy regimes.

\subsection*{Acknowledgements}

We thank Ben Bullock for useful discussions.
This work is supported in part by the
Ministero dell' Universit\`a e della Ricerca Scientifica, the UK PPARC,
and   the EC Programme
``Human Capital and Mobility'', Network ``Physics at High Energy
Colliders'', contract CHRX-CT93-0357, DG 12 COMA (SM).
KO thanks Trinity College and the Committee of Vice-Chancellors and Principals
of the Universities of the United Kingdom for financial support.

\goodbreak


\section*{Table Captions}

\begin{itemize}

\item[{[I]}] Cross section in femtobarns for processes (1)--(2),
in the channels: 
(a) ${\mathrm{jj}} {\mathrm{jj}}$,
(b) ${\mathrm{jj}} \tau\nu_\tau$,
(c) $\tau\nu_\tau \tau\nu_\tau$;
for three different values of CM energy, before (upper section)
and after (lower section) the following cuts:
(a) and
(b) $|\cos\theta_{\mathrm{B}}|<0.60$ (for all $\sqrt s$), plus
$\cos\theta_{\mathrm{ff}}$$<-0.88(-0.80)[-0.72]$ 
and $p_{\mathrm{T}}<24(34)[40]$ GeV, in
correspondence of  $\sqrt s=205(210)[215]$ GeV;
(c) $\cos\theta_{\mathrm{ff}}$$>-0.50$ and $p_{\mathrm{T}}>50$ GeV, for 
$\sqrt s=205,210,215$ GeV.
The errors on the cross sections are as given by {\tt VEGAS}.

\end{itemize}

\section*{Figure Captions}

\begin{itemize}

\item[{[1]}] Differential distribution in $\cos \theta_{\mathrm{B}}$
(see the text for the definition depending on the final state),
for processes (1)--(2), in the channels:
(a) ${\mathrm{jj}} {\mathrm{jj}}$,
(b) ${\mathrm{jj}} \tau\nu_\tau$,
(c) $\tau\nu_\tau \tau\nu_\tau$;
for three different values of CM energy, before any cut.
Spectra are normalised to one. 
Continuous line: $W^+W^-$; dashed line: $H^+H^-$.

\item[{[2]}] Differential distribution in $\cos \theta_{\mathrm{ff}}$
(see the text for the definition depending on the final state),
$W$ and $H$ bosons,
for processes (1)--(2), in the channels:
(a) ${\mathrm{jj}} {\mathrm{jj}}$,
(b) ${\mathrm{jj}} \tau\nu_\tau$,
(c) $\tau\nu_\tau \tau\nu_\tau$;
for three different values of CM energy, before any cut.
Spectra are normalised to one. 
Continuous line: $W^+W^-$; dashed line: $H^+H^-$.

\item[{[3]}] Differential distribution in $p_{\mathrm{T}}$
(see the text for the definition depending on the final state),
for processes (1)--(2), in the channels:
(a) ${\mathrm{jj}} {\mathrm{jj}}$,
(b) ${\mathrm{jj}} \tau\nu_\tau$,
(c) $\tau\nu_\tau \tau\nu_\tau$;
for three different values of CM energy, before any cut.
Spectra are normalised to one. 
Continuous line: $W^+W^-$; dashed line: $H^+H^-$.

\item[{[4]}] Differential distribution in $M_{\mathrm{ff}}$
(see the text for the definition depending on the final state),
for processes (1)--(2), in the channels:
(a) ${\mathrm{jj}} {\mathrm{jj}}$,
(b) ${\mathrm{jj}} \tau\nu_\tau$,
(c) $\tau\nu_\tau \tau\nu_\tau$;
for three different values of CM energy, 
after the following cuts:
(a) and
(b) $|\cos\theta_{\mathrm{B}}|<0.60$ (for all $\sqrt s$), plus
$\cos\theta_{\mathrm{ff}}$$<-0.88(-0.80)[-0.72]$ 
and $p_{\mathrm{T}}<24(34)[40]$ GeV, in
correspondence of  $\sqrt s=205(210)[215]$ GeV;
(c) $\cos\theta_{\mathrm{ff}}$$>-0.50$ and $p_{\mathrm{T}}>50$ GeV, for 
$\sqrt s=205,210,215$ GeV.
Spectra are normalised to the total cross sections. 
Continuous line: $W^+W^-$; dashed line: $H^+H^-$.

\end{itemize}
\vfill
\clearpage

\
\vskip8.0cm
\begin{table}
\begin{center}
\begin{tabular}{|c|c|c|}
\hline
\multicolumn{3}{|c|}
{\rule[0cm]{0cm}{0cm}
$\sigma$ (fb)}
\\ \hline
\rule[0cm]{0cm}{0cm}
$\sqrt s$ (GeV) & $W^+W^-\rightarrow \mathrm{jj} \mathrm{jj}$
                        & $H^+H^-\rightarrow \mathrm{jj} \mathrm{jj}$
\\ \hline\hline

\rule[0cm]{0cm}{0cm}
$205$ & $6904.\pm10.$ & $(1212.11\pm0.31)\times10^{-4}$ \\ \hline
\rule[0cm]{0cm}{0cm}
$210$ & $6850.\pm11.$ & $(2967.12\pm0.77)\times10^{-4}$ \\ \hline
\rule[0cm]{0cm}{0cm}
$215$ & $6777.\pm11.$ & $(4852.04\pm1.30)\times10^{-4}$ \\ \hline
\multicolumn{3}{|c|}
{\rule[0cm]{0cm}{0cm}
no kinematic cuts}
 \\ \hline\hline

\rule[0cm]{0cm}{0cm}
$205$ & $278.9\pm1.1$ & $(9.5164\pm0.0034)\times10^{-2}$ \\ \hline
\rule[0cm]{0cm}{0cm}
$210$ & $458.4\pm1.8$ & $(22.9843\pm0.0090)\times10^{-2}$ \\ \hline
\rule[0cm]{0cm}{0cm}
$215$ & $616.1\pm2.0$ & $(33.112\pm0.010)\times10^{-2}$ \\ \hline
\multicolumn{3}{|c|}
{\rule[0cm]{0cm}{0cm}
after kinematic cuts}
 \\ \hline\hline

\multicolumn{3}{|c|}
{\rule[0cm]{0cm}{0cm}
$M_W\approx80$ GeV
\qquad\qquad
$\tan\beta=1.5$
\qquad\qquad
$M_H\approx100$ GeV}
 \\ \hline
\multicolumn{3}{c}
{\rule{0cm}{1.0cm}
{\Large Tab. Ia}}  \\
\multicolumn{3}{c}
{\rule{0cm}{0cm}}

\end{tabular}
\end{center}
\end{table}
\vfill
\clearpage

\
\vskip8.0cm
\begin{table}
\begin{center}
\begin{tabular}{|c|c|c|}
\hline
\multicolumn{3}{|c|}
{\rule[0cm]{0cm}{0cm}
$\sigma$ (fb)}
\\ \hline
\rule[0cm]{0cm}{0cm}
$\sqrt s$ (GeV) & $W^+W^-\rightarrow {\mathrm{jj}} \tau\nu_\tau$
                        & $H^+H^-\rightarrow {\mathrm{jj}} \tau\nu_\tau$
\\ \hline\hline

\rule[0cm]{0cm}{0cm}
$205$ & $2300.79\pm3.34$ & $1.6207\pm0.0042$ \\ \hline
\rule[0cm]{0cm}{0cm}
$210$ & $2282.68\pm3.58$ & $3.967\pm0.010$ \\ \hline
\rule[0cm]{0cm}{0cm}
$215$ & $2258.24\pm3.82$ & $6.488\pm0.017$ \\ \hline
\multicolumn{3}{|c|}
{\rule[0cm]{0cm}{0cm}
no kinematic cuts}
 \\ \hline\hline

\rule[0cm]{0cm}{0cm}
$205$ & $2.140\pm0.014$ & $1.2835\pm0.0044$ \\ \hline
\rule[0cm]{0cm}{0cm}
$210$ & $4.651\pm0.024$ & $3.1427\pm0.0051$ \\ \hline
\rule[0cm]{0cm}{0cm}
$215$ & $6.714\pm0.031$ & $5.1384\pm0.0085$ \\ \hline
\multicolumn{3}{|c|}
{\rule[0cm]{0cm}{0cm}
after kinematic cuts}
 \\ \hline\hline

\multicolumn{3}{|c|}
{\rule[0cm]{0cm}{0cm}
$M_W\approx80$ GeV
\qquad\qquad
$\tan\beta=1.5$
\qquad\qquad
$M_H\approx100$ GeV}
 \\ \hline
\multicolumn{3}{c}
{\rule{0cm}{1.0cm}
{\Large Tab. Ib}}  \\
\multicolumn{3}{c}
{\rule{0cm}{0cm}}

\end{tabular}
\end{center}
\end{table}
\vfill
\clearpage

\
\vskip8.0cm
\begin{table}
\begin{center}
\begin{tabular}{|c|c|c|}
\hline
\multicolumn{3}{|c|}
{\rule[0cm]{0cm}{0cm}
$\sigma$ (fb)}
\\ \hline
\rule[0cm]{0cm}{0cm}
$\sqrt s$ (GeV) & $W^+W^-\rightarrow \tau\nu_\tau \tau\nu_\tau$
                        & $H^+H^-\rightarrow \tau\nu_\tau \tau\nu_\tau$
\\ \hline\hline

\rule[0cm]{0cm}{0cm}
$205$ & $191.68\pm0.28$ & $5.4177\pm0.0014$ \\ \hline
\rule[0cm]{0cm}{0cm}
$210$ & $190.17\pm0.30$ & $13.2620\pm0.0034$ \\ \hline
\rule[0cm]{0cm}{0cm}
$215$ & $188.14\pm0.32$ & $21.6870\pm0.0056$ \\ \hline
\multicolumn{3}{|c|}
{\rule[0cm]{0cm}{0cm}
no kinematic cuts}
 \\ \hline\hline

\rule[0cm]{0cm}{0cm}
$205$ & $35.481\pm0.085$ & $2.8796\pm0.0018$ \\ \hline
\rule[0cm]{0cm}{0cm}
$210$ & $33.563\pm0.084$ & $6.9991\pm0.0044$ \\ \hline
\rule[0cm]{0cm}{0cm}
$215$ & $31.555\pm0.082$ & $11.3539\pm0.0073$ \\ \hline
\multicolumn{3}{|c|}
{\rule[0cm]{0cm}{0cm}
after kinematic cuts}
 \\ \hline\hline

\multicolumn{3}{|c|}
{\rule[0cm]{0cm}{0cm}
$M_W\approx80$ GeV
\qquad\qquad
$\tan\beta=1.5$
\qquad\qquad
$M_H\approx100$ GeV}
 \\ \hline
\multicolumn{3}{c}
{\rule{0cm}{1.0cm}
{\Large Tab. Ic}}  \\
\multicolumn{3}{c}
{\rule{0cm}{0cm}}

\end{tabular}
\end{center}
\end{table}
\vfill
\clearpage
\begin{figure}[p]
\centerline{\epsfig{figure=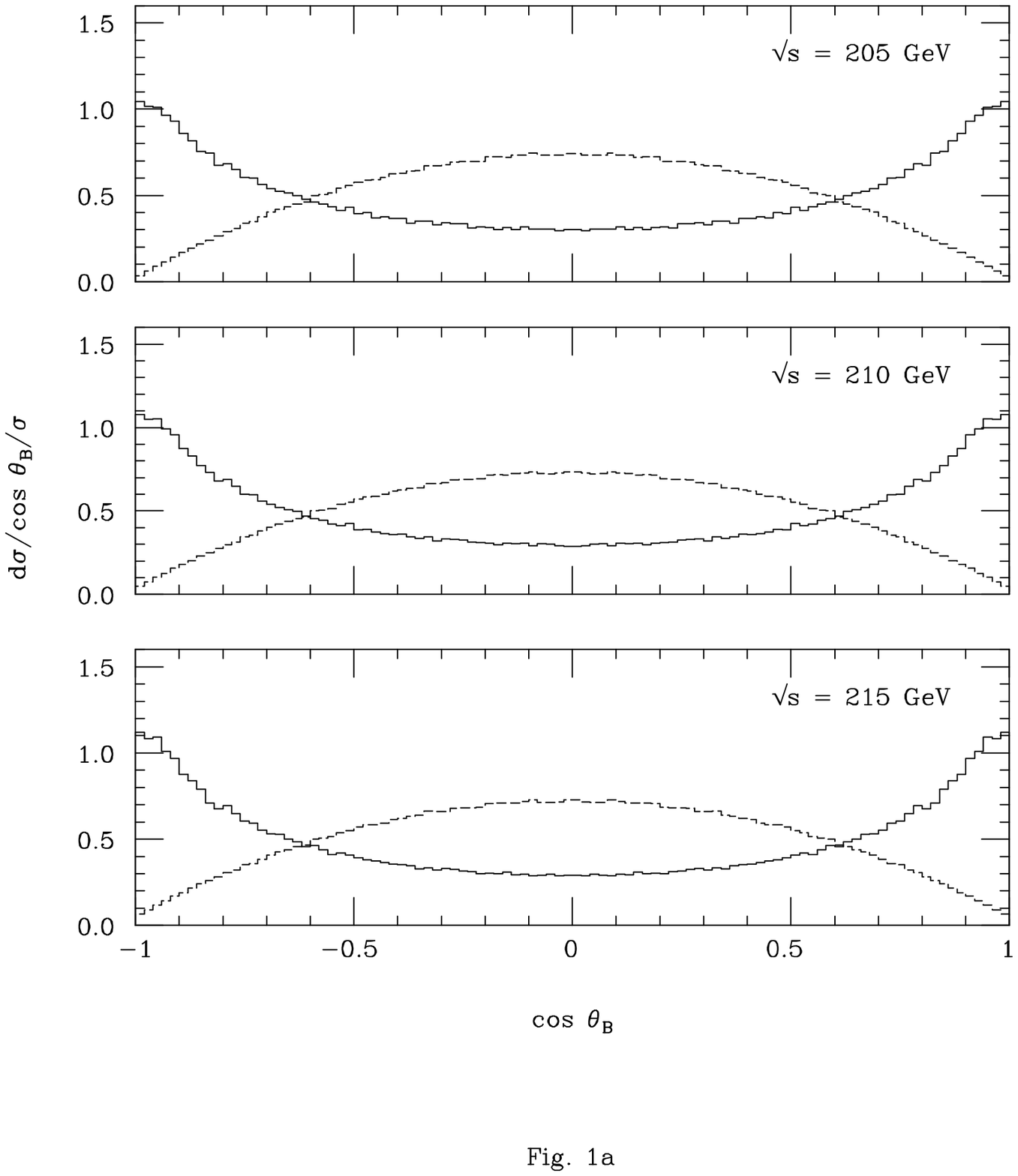,height=20cm}}  
\vspace*{2cm}
\end{figure}
\vfill
\clearpage

\begin{figure}[p]
\centerline{\epsfig{figure=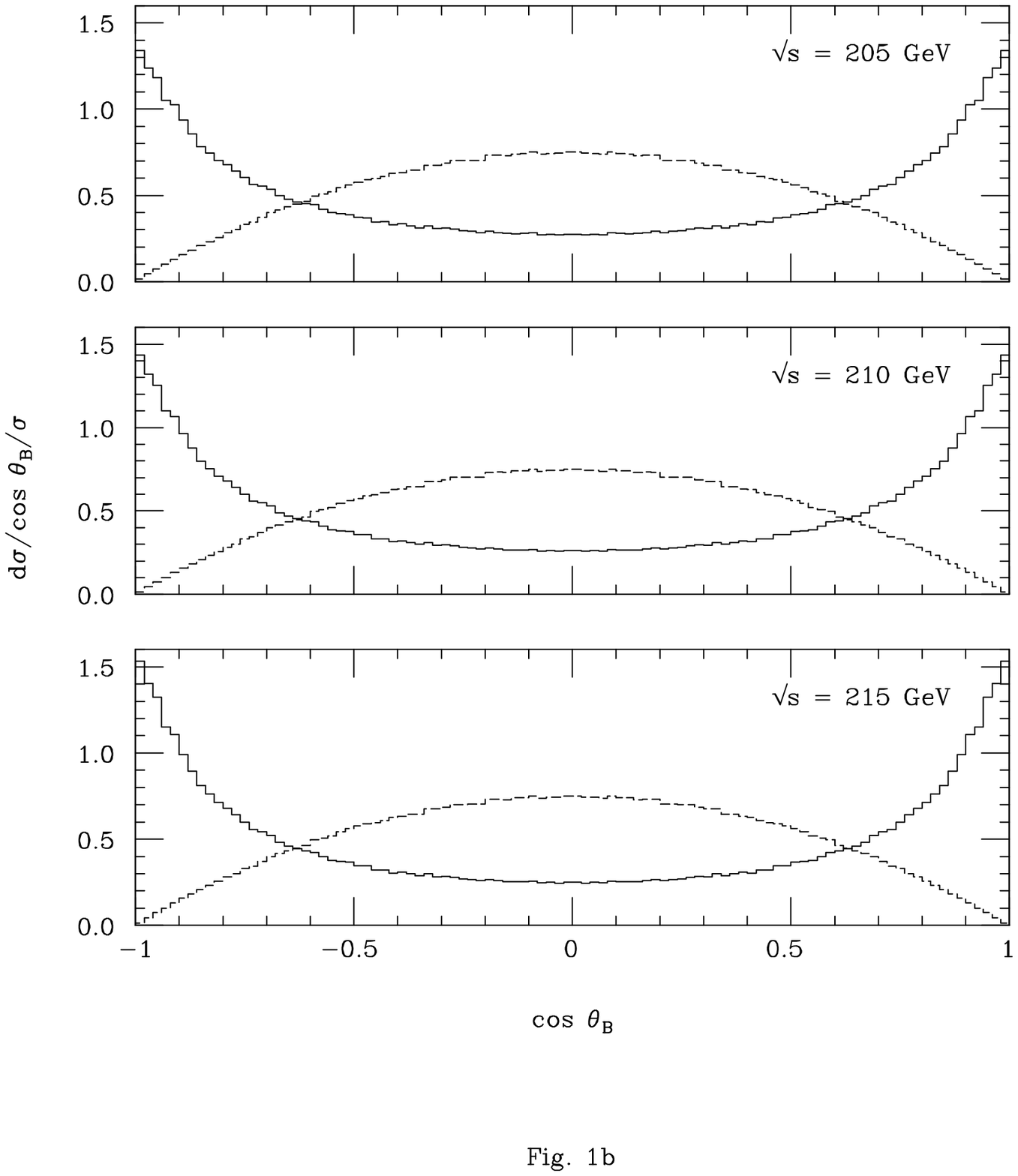,height=20cm}}  
\vspace*{2cm}
\end{figure}
\vfill
\clearpage
\begin{figure}[p]
\centerline{\epsfig{figure=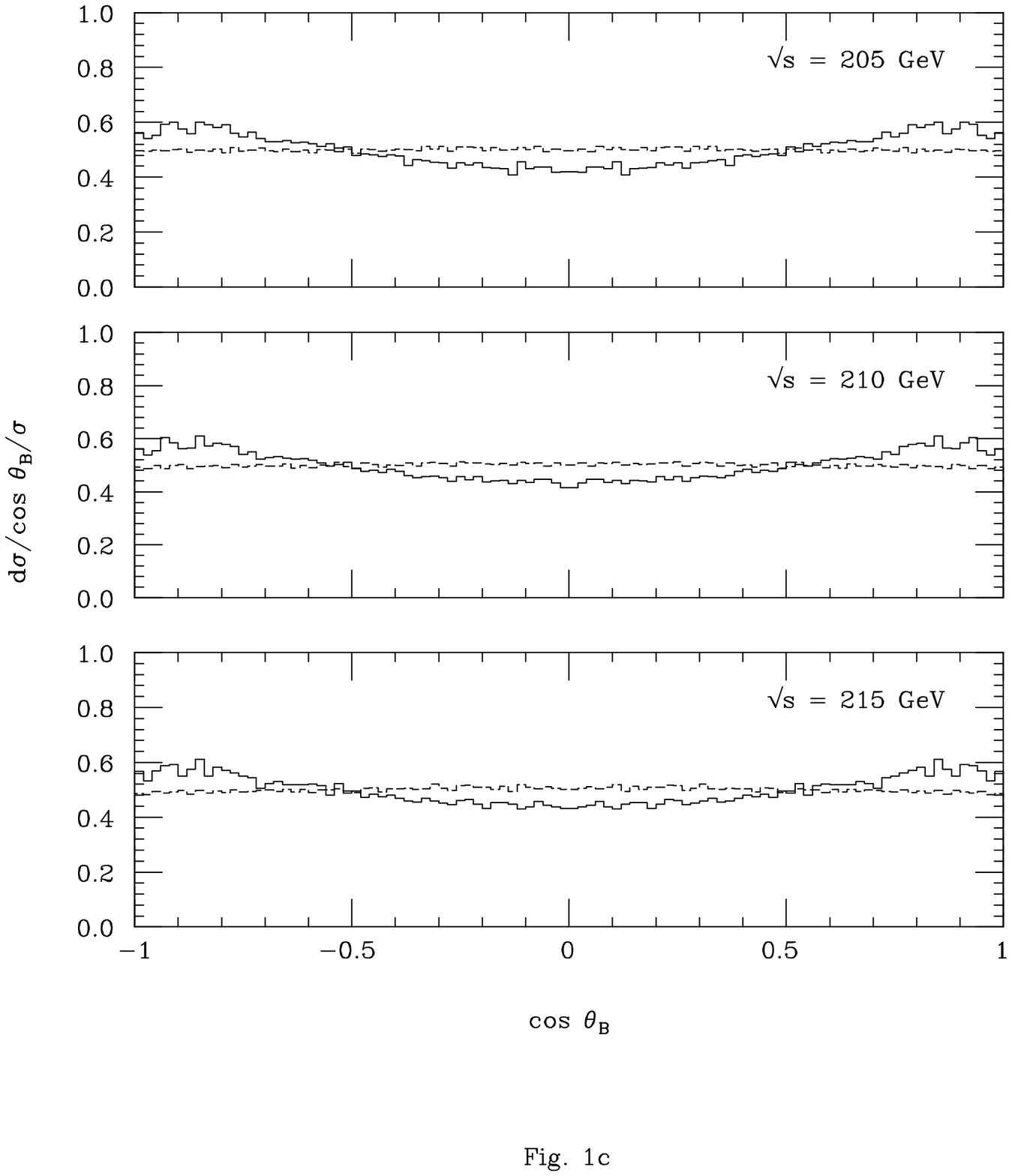,height=20cm}}  
\vspace*{2cm}
\end{figure}
\vfill
\clearpage
\begin{figure}[p]
\centerline{\epsfig{figure=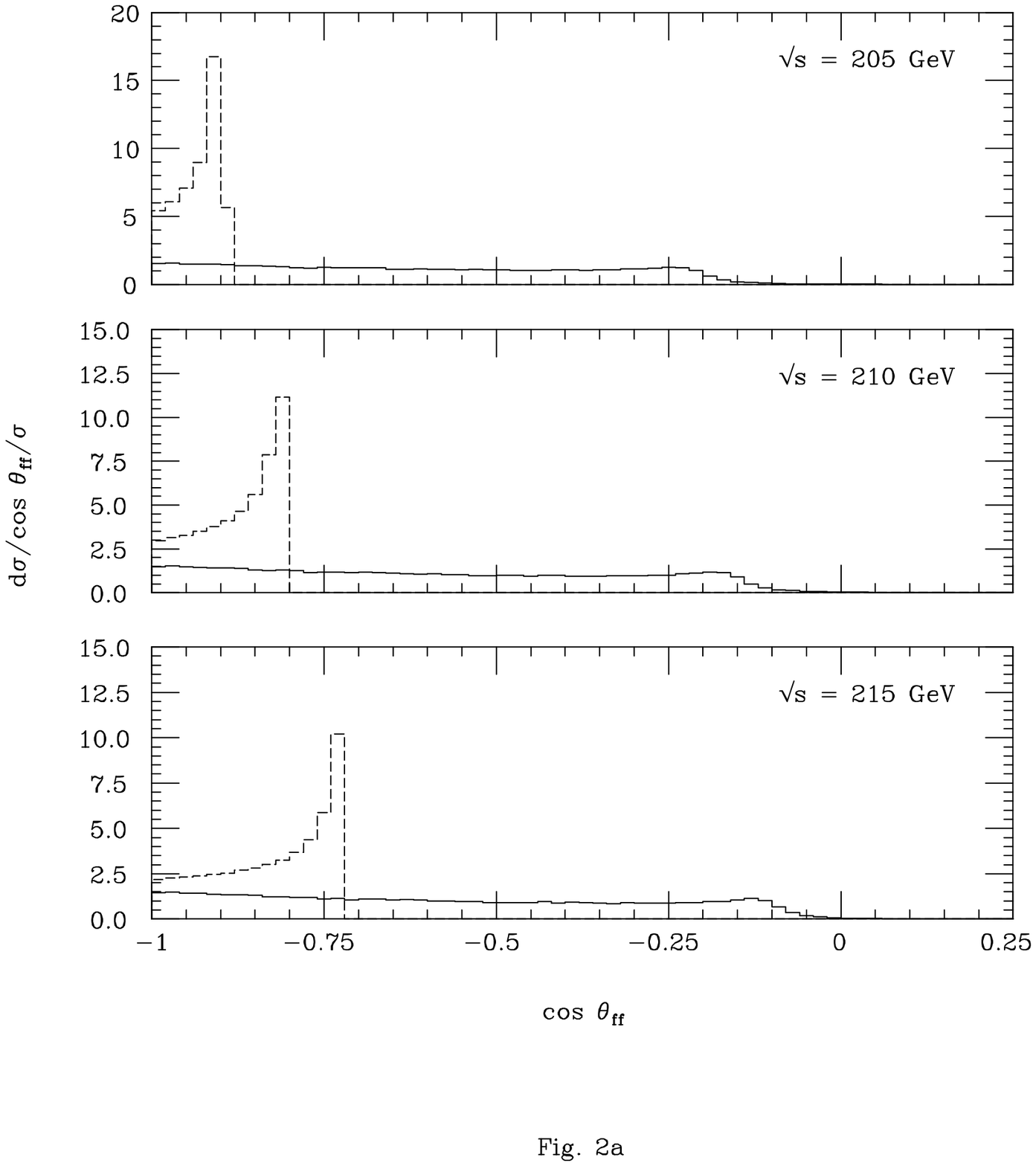,height=20cm}}  
\vspace*{2cm}
\end{figure}
\vfill
\clearpage
\begin{figure}[p]
\centerline{\epsfig{figure=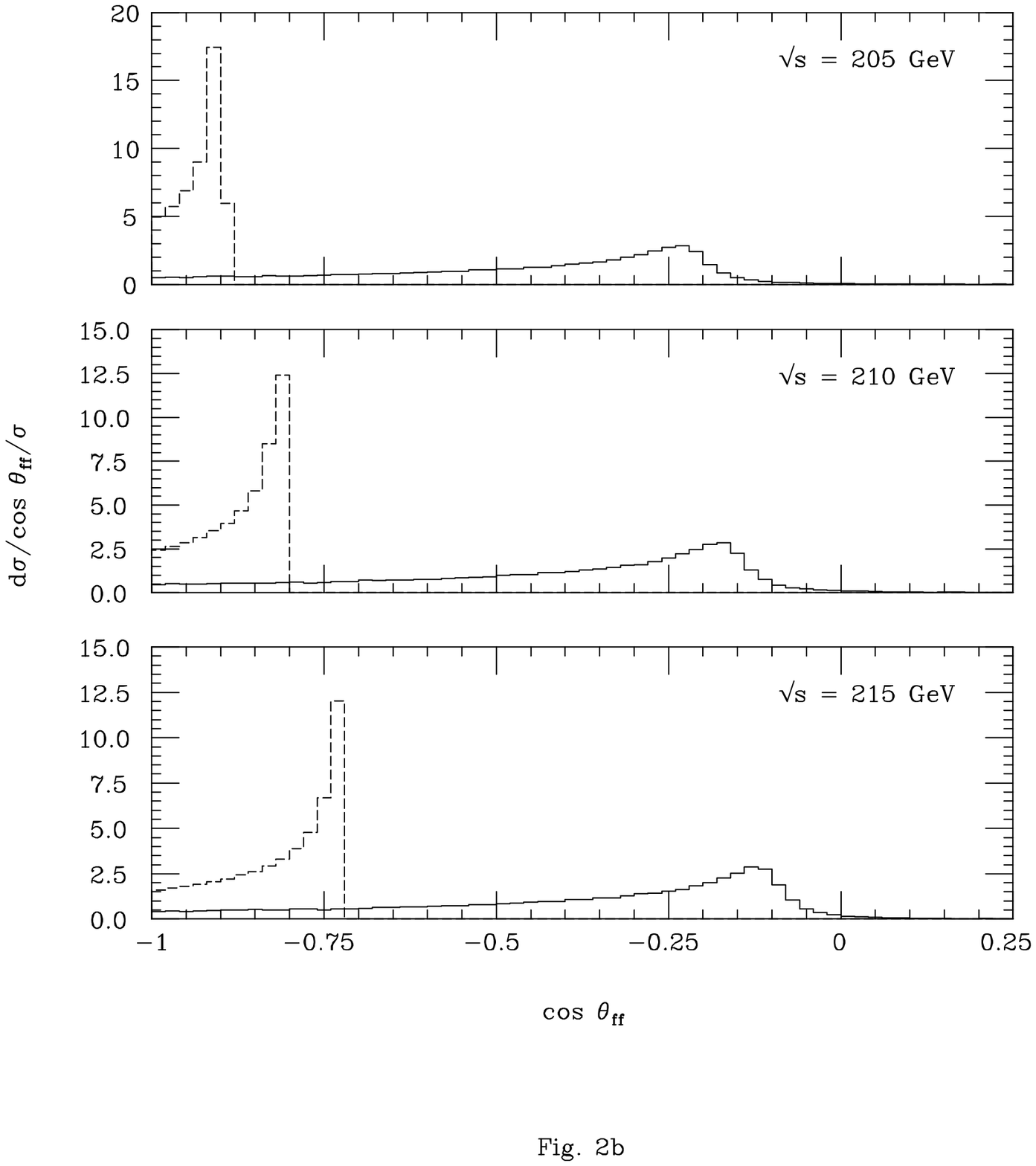,height=20cm}}  
\vspace*{2cm}
\end{figure}
\vfill
\clearpage
\begin{figure}[p]
\centerline{\epsfig{figure=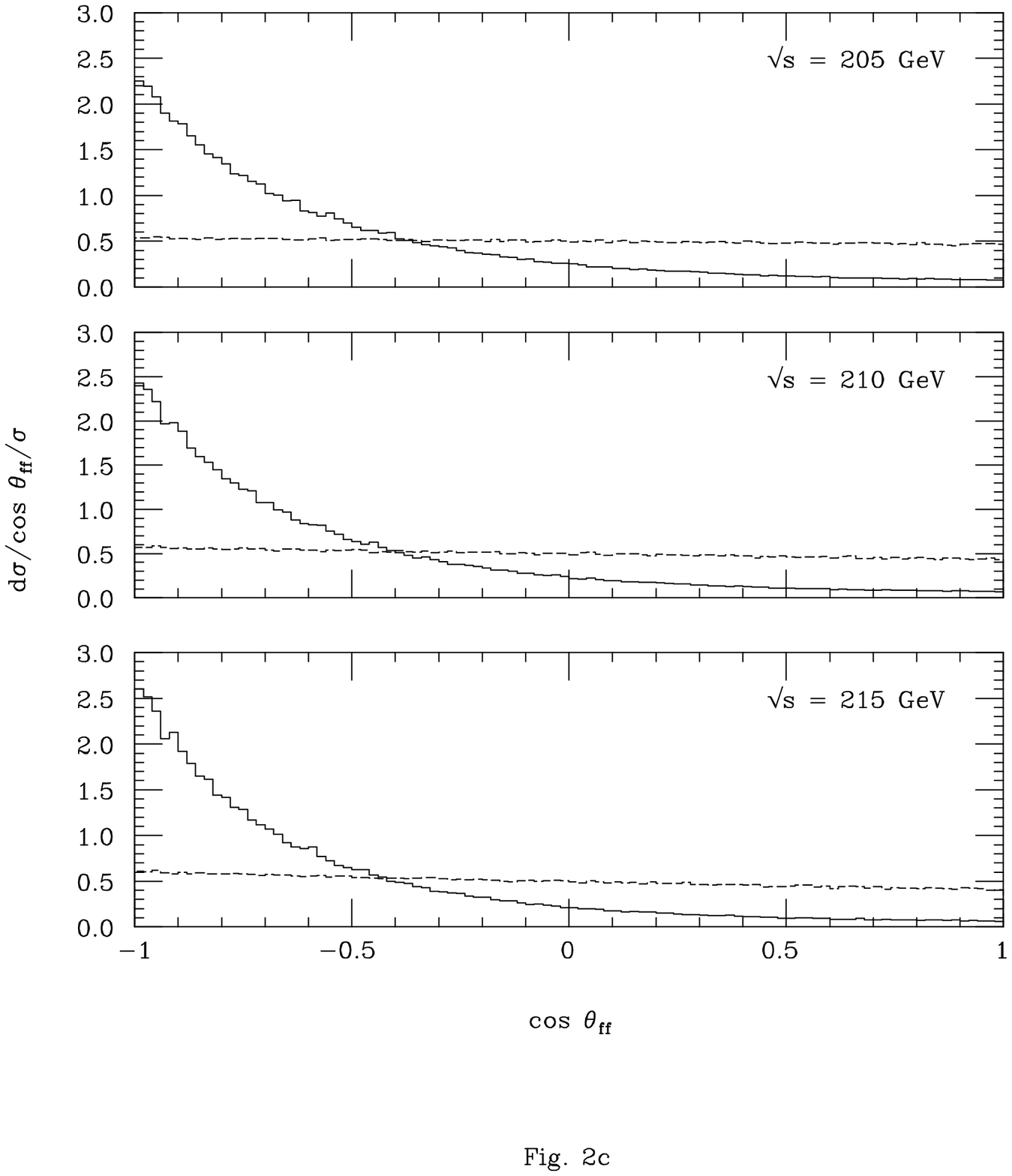,height=20cm}}  
\vspace*{2cm}
\end{figure}
\vfill
\clearpage
\begin{figure}[p]
\centerline{\epsfig{figure=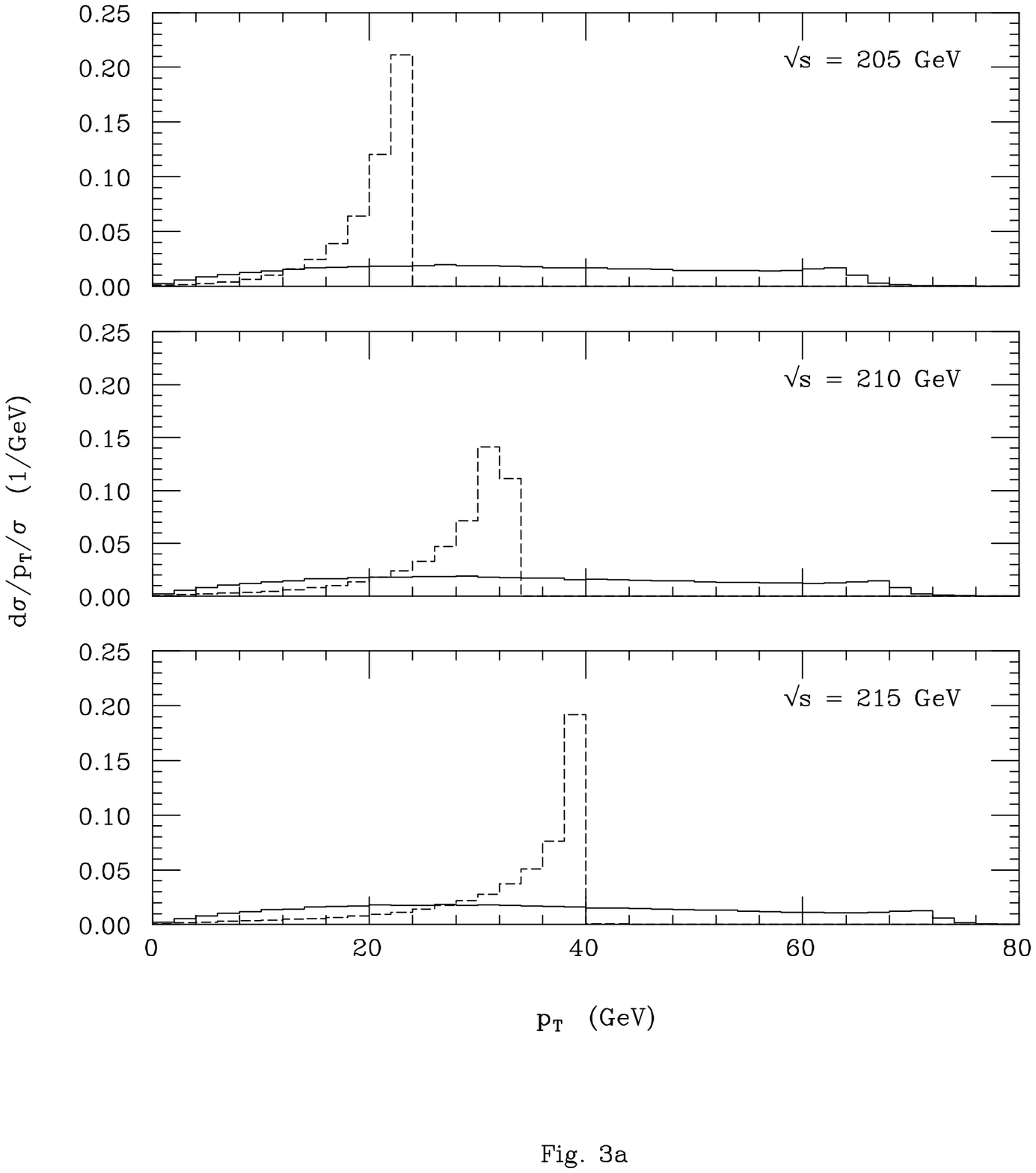,height=20cm}}  
\vspace*{2cm}
\end{figure}
\vfill
\clearpage
\begin{figure}[p]
\centerline{\epsfig{figure=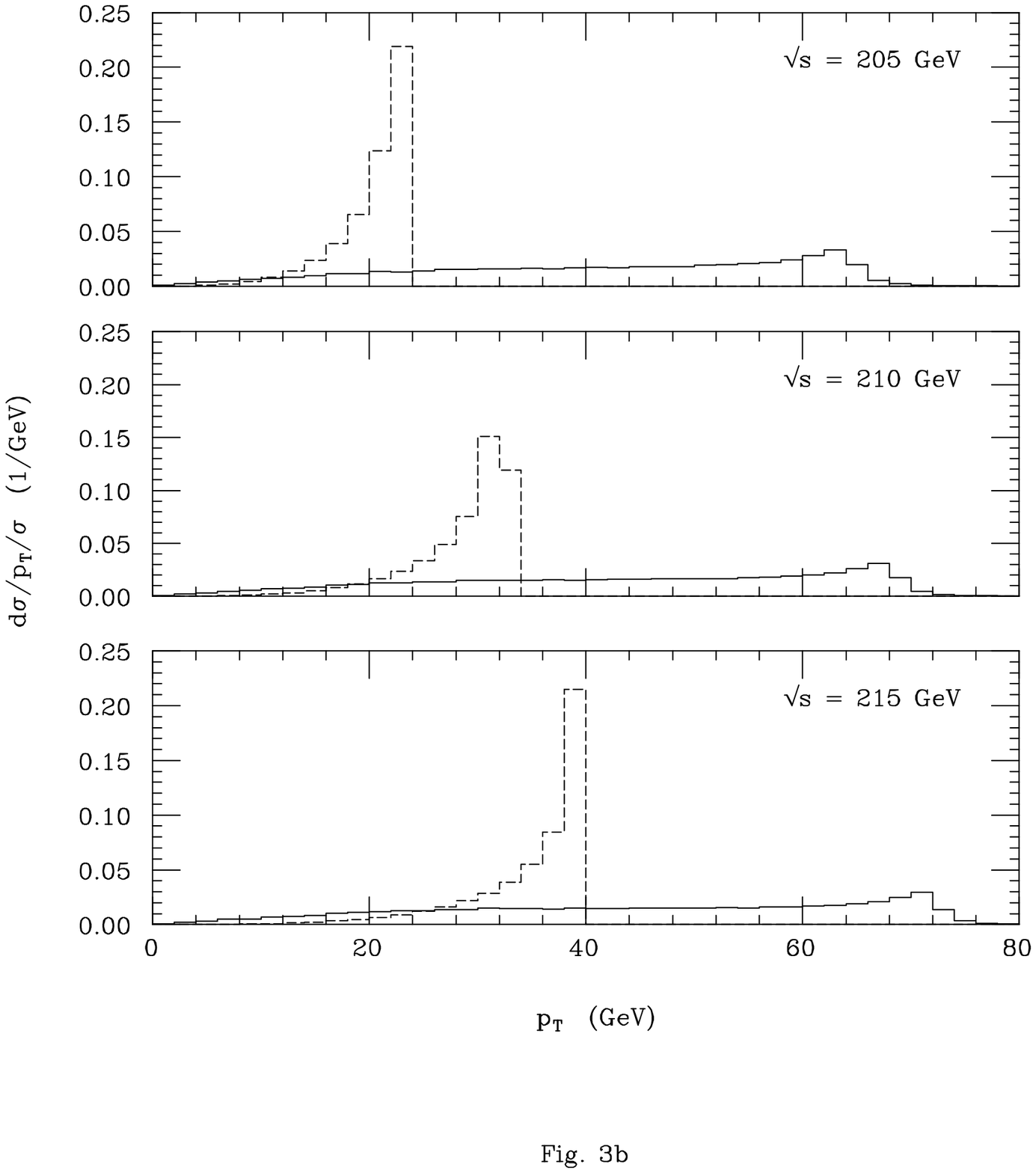,height=20cm}}  
\vspace*{2cm}
\end{figure}
\vfill
\clearpage
\begin{figure}[p]
\centerline{\epsfig{figure=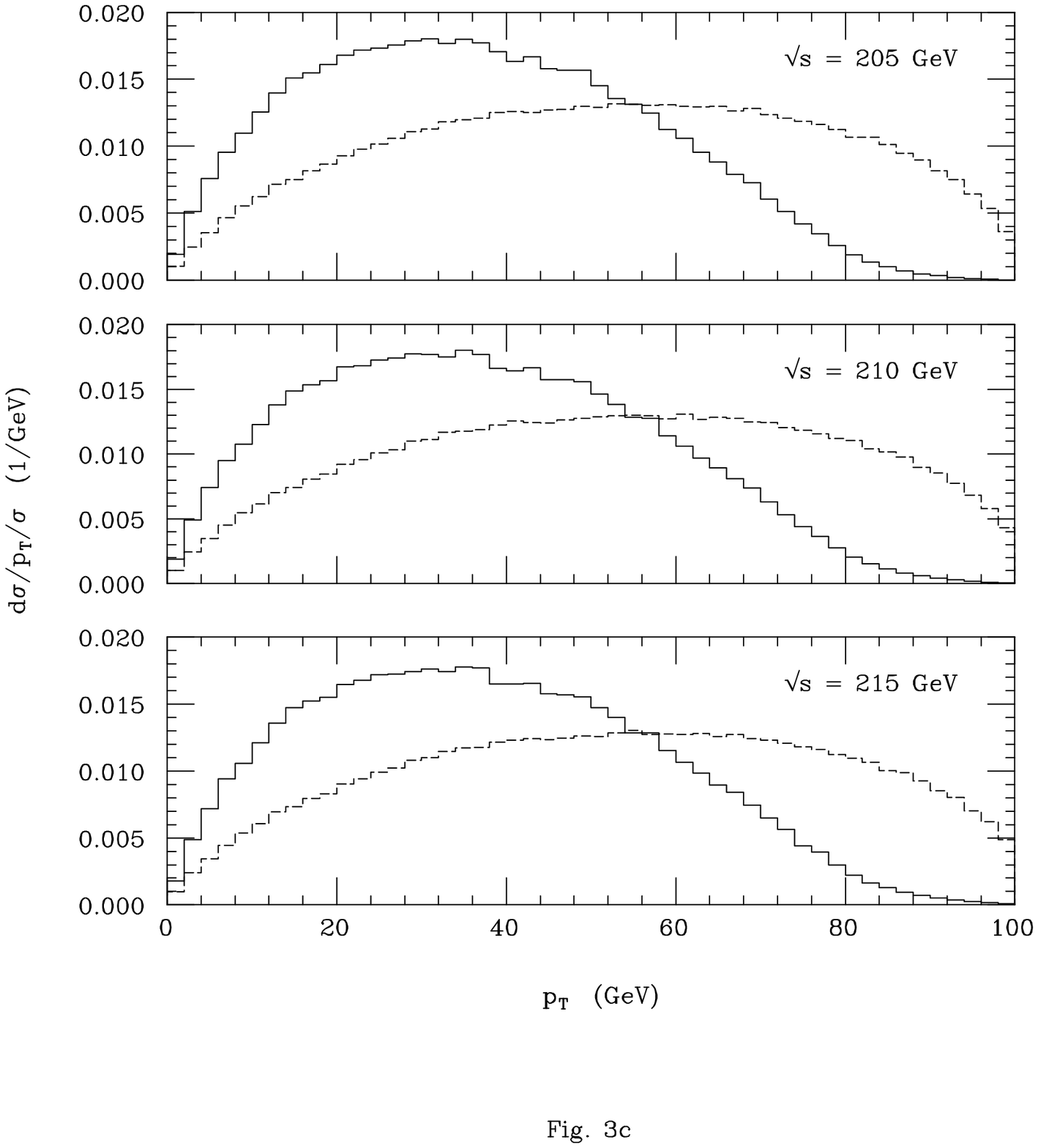,height=20cm}}  
\vspace*{2cm}
\end{figure}
\vfill
\clearpage
\begin{figure}[p]
\centerline{\epsfig{figure=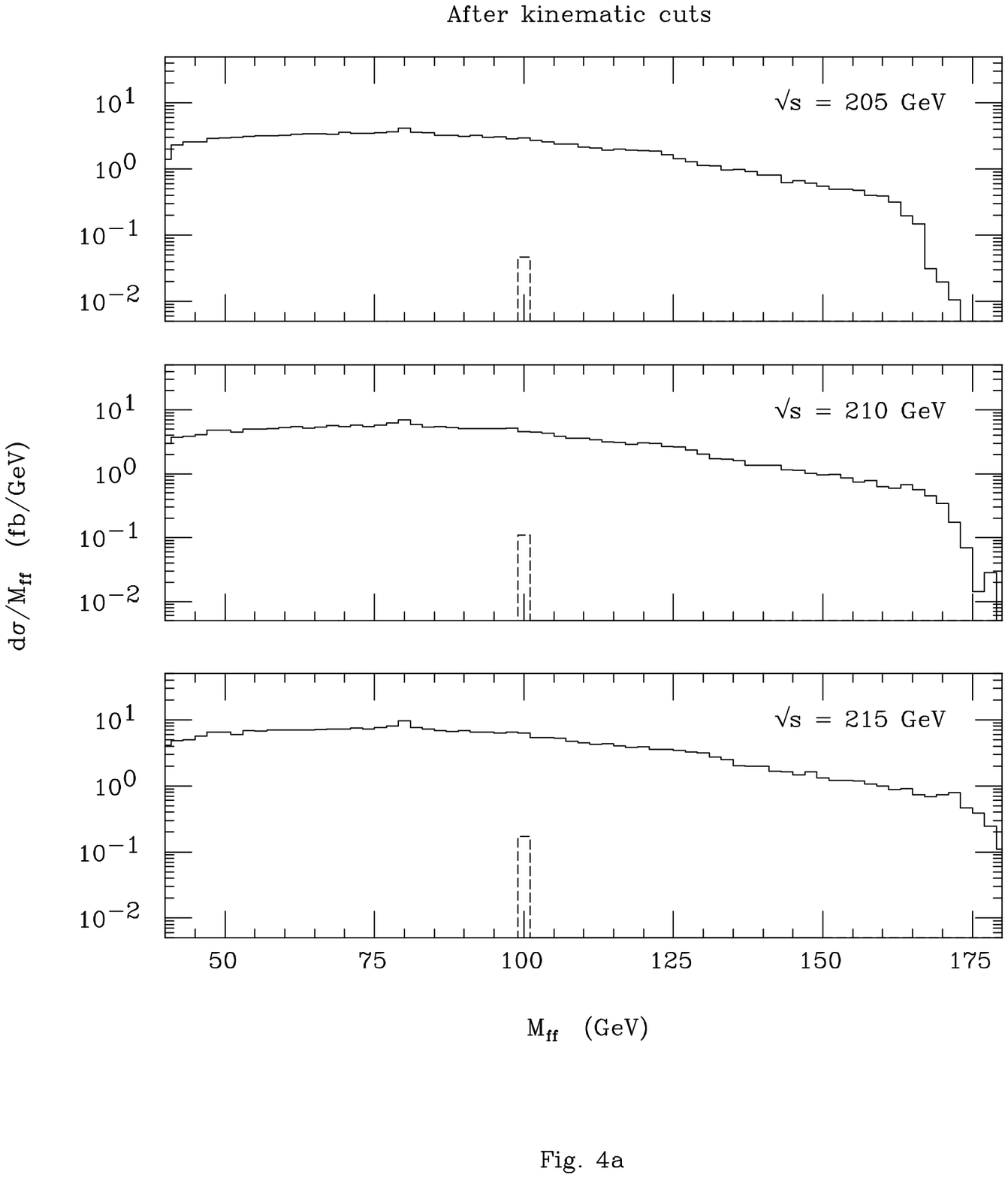,height=20cm}}  
\vspace*{2cm}
\end{figure}
\vfill
\clearpage
\begin{figure}[p]
\centerline{\epsfig{figure=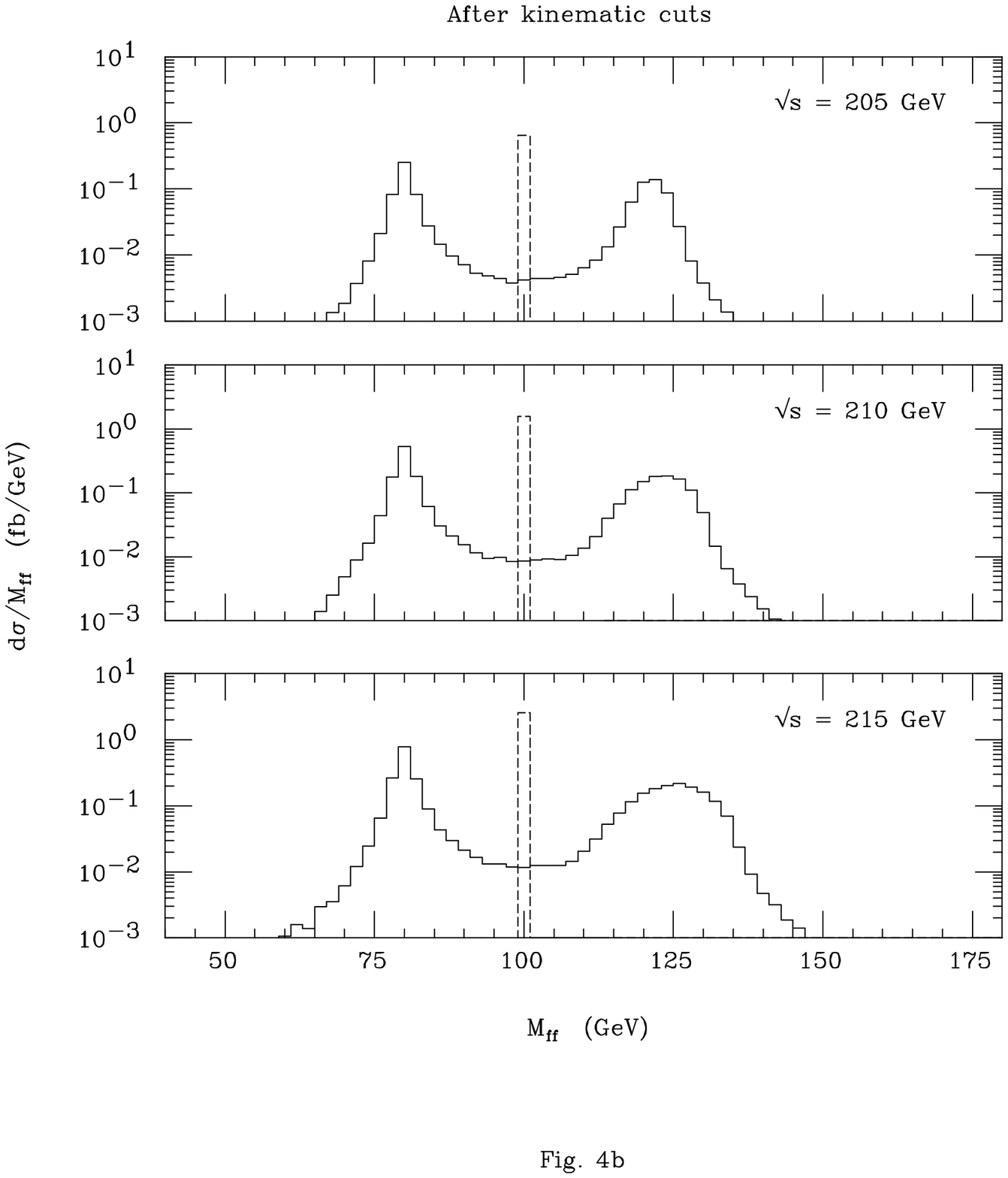,height=20cm}}  
\vspace*{2cm}
\end{figure}
\vfill
\clearpage
\begin{figure}[p]
\centerline{\epsfig{figure=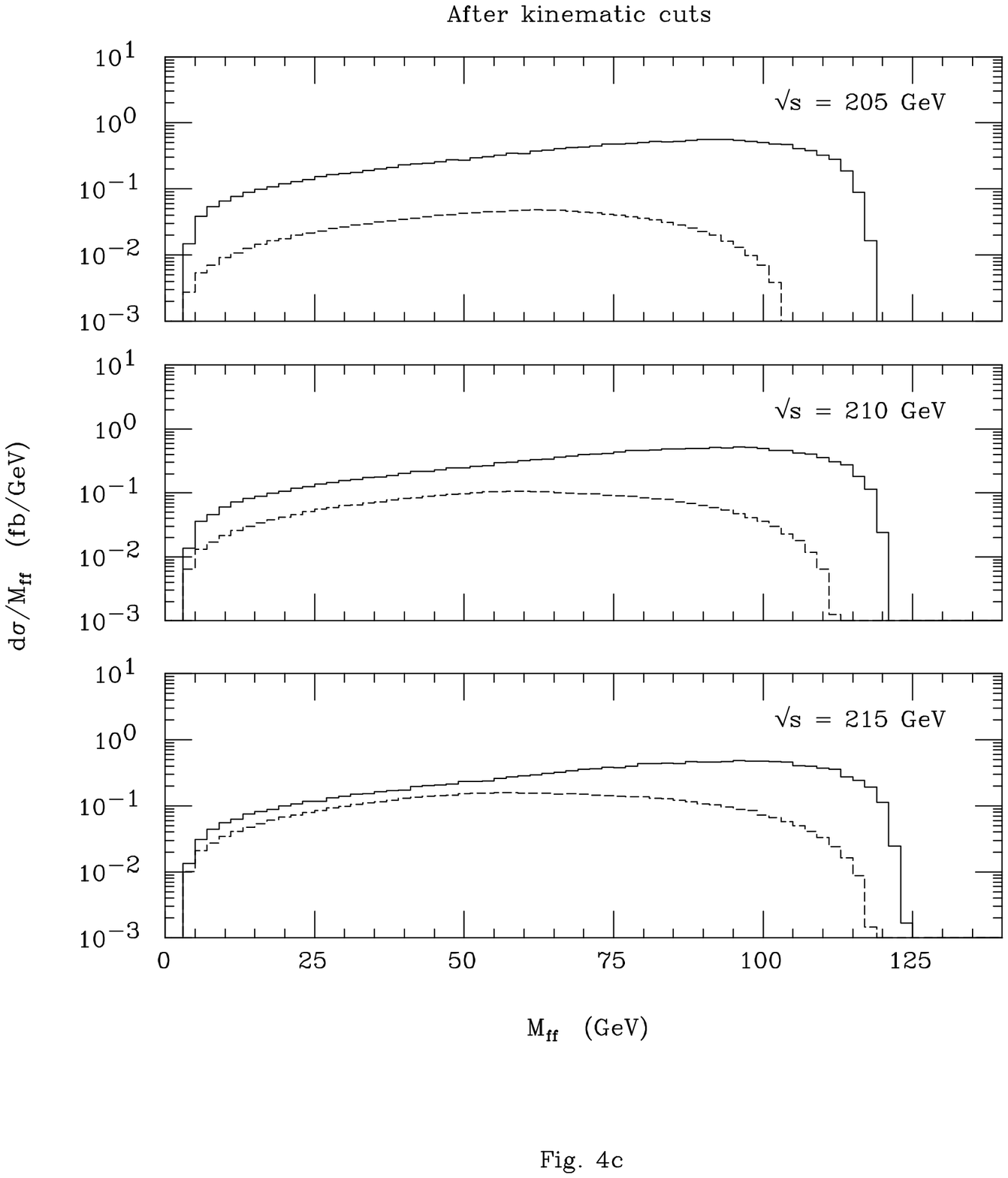,height=20cm}}  
\vspace*{2cm}
\end{figure}
\vfill
\clearpage

\end{document}